\documentclass[traditabstract]{aa} 
\usepackage{graphicx}
\usepackage{txfonts}
\usepackage{natbib}
\usepackage{textcomp}
\usepackage{longtable}
\usepackage{booktabs}
\usepackage{lscape}
\usepackage{xspace}
\usepackage{tabularx}
\begin{document}
\title{Size and disk-like shape of the broad-line region of ESO399-IG20}


	\author{
          Francisco Pozo Nu\~nez 
          \inst{1} 
          \and
          Christian Westhues
          \inst{1}
          \and
          Michael Ramolla          
          \inst{1}
          \and
          Christoph Bruckmann
          \inst{1}
          \and 
          Martin Haas
          \inst{1}
          \and 
          Rolf Chini
          \inst{1,2}
          \and
          Katrien Steenbrugge
          \inst{2,3}		
          \and
          Roland Lemke
          \inst{1}
          \and
          Miguel Murphy
          \inst{2}}
	\institute{
          Astronomisches Institut, Ruhr--Universit\"{a}t Bochum,
	  Universit\"{a}tsstra{\ss}e 150, 44801 Bochum, Germany
	  \and
          Instituto de Astronomia, Universidad Cat\'{o}lica del
          Norte, Avenida Angamos 0610, Casilla
          1280 Antofagasta, Chile
        \and
          Department of Physics, 
          University of Oxford, 
          Keble Road,
          Oxford OX1 3RH, UK\\}
        

	\date{Received January 6, 2013; Accepted February 5, 2013}
        
	\abstract{We present photometric reverberation mapping of 
          the narrow-line Seyfert 1 galaxy \object{ESO399-IG20} 
          performed with the robotic 15 cm telescope VYSOS-6 at the Cerro Armazones Observatory. 
          Through the combination of broad- and narrow-band filters we determine 
          the size of the broad-line emitting region (BLR) by measuring 
          the time delay between the variability of the continuum and the H$\alpha$ emission line. 
          We use the flux variation gradient method to separate the host galaxy 
          contribution from that of the active galactic nucleus (AGN), and to calculate 
          the $5100\AA$ luminosity $L_{\rm{AGN}}$ of the AGN. 
          Both measurements permit us to derive the position of ESO399-IG20 
          in the BLR size -- AGN luminosity 
          $R_{\rm{BLR}} \propto L_{\rm{AGN}}^{0.5}$ diagram. 
          We infer the basic geometry of the BLR through 
          modelling of the light curves. 
          The pronounced sharp variability patterns in both the continuum and the emission 
          line light curves allow us to reject a spherical BLR geometry. 
          The light curves are best fitted by a disk-like BLR seen nearly face-on with an 
          inclination angle of 6$^{\circ}\pm$3$^{\circ}$ and with an extension from 16 to 20 light days.
        }
        
		\maketitle
%

\section{Introduction}

The physical properties of the broad line region (BLR) in 
active galactic nuclei (AGN) have been extensively studied
during the last three decades. Consensus has been reached that 
the luminosity variations of the continuum 
emitting region of the hot accretion disk produce variations of the 
broad emission lines with a delay due to the light travel time across the BLR.
To date, the only method independent on spatial resolution is reverberation mapping 
(\citealt{1982ApJ...255..419B}; \citealt{1993PASP..105..247P}), in which one measures 
the time delay or "echo" between the continuum variability and the variability observed 
in the broad emission lines. This method has been 
used successfully for determining the BLR size, kinematics and black hole mass ($M_{BH}$)
of several Type-1 AGNs. RM provides basic $M_{BH}$ estimates through the virial product 
${M_{BH}} = f \cdot R_{BLR} \cdot \sigma_{V}^2/{G}$, where $G$ is the 
gravitational constant, $R_{BLR}=c \cdot \tau$ is the BLR size, $\sigma_{V}$ is the 
emission-line velocity dispersion of the BLR gas and the factor $f$ depends on 
the -- so far unknown -- geometry 
and kinematics of the BLR (\citealt{2004ApJ...613..682P} and references therein). 
Furthermore, as a widely used extrapolation of locally obtained RM results 
to higher redshift data,  $M_{BH}$ has be 
estimated using the BLR Size-Luminosity $R_{\rm{BLR}} \propto L_{\rm{AGN}}^{0.5}$ 
relationship (\citealt{1991ApJ...370L..61K}; \citealt{1996ApJ...471L..75K}, \citealt{2000ApJ...533..631K}; \citealt{1999ApJ...526..579W}; 
\citealt{2008ApJ...673..703M}; \citealt{2011nlsg.confE..38V}). 
However, a reliable $M_{BH}$ estimation from the luminosity requires 
a considerable reduction of the $R - L$ dispersion present.

As the size of the BLR 
ranges from a few to several hundred light days, it is necessary to perform observational 
monitoring campaigns ranging from months to years in order to sufficiently sample 
the time domain of the echo. Most observational RM campaigns 
have used spectroscopic monitoring of the sources, which however
requires large amounts of telescope time.

Recently, \citet{2011A&A...535A..73H} proposed photometric reverberation mapping (PRM) 
as an efficient method to determine the BLR size
through the use of broad band filters 
to trace the AGN continuum variations and narrow-band filters to catch the BLR emission 
line response. 
Because the narrow-band collects both the emission line flux and the underlying continuum, 
the challenge is to extract the pure emission line light curve.
In the dedicated case study of 3C120, \citet{2012A&A...545A..84P} demonstrated that 
PRM reaches an accuracy similar to spectroscopic RM (\citealt{2012arXiv1206.6523G}). 

Another efficient approach 
to determine the BLR size has been proposed by \citet{2012ApJ...747...62C},
through the analysis of the difference between the cross-correlation (CCF) and 
the auto-correlation (ACF) functions for suitably chosen broad band filters, 
one filter covering a bright emission line and one covering only the continuum. 
While this method does not place a strict requirement on the 
object's redshift such that the emission line falls into the narrow-band filter 
(as in the case proposed by \citealt{2011A&A...535A..73H}), 
the use of broad band filters only (as proposed by \citealt{2012ApJ...747...62C}) 
limits the application of PRM to cases with a sufficiently strong emission line contribution 
in the 
respective filter used. Despite of this handicap, this method has been successfully applied to determine 
the BLR size for one low-luminosity AGN NGC4395 (\citealt{2012ApJ...756...73E}) and one high-$z$ luminous MACHO quasar
(\citealt{2012ApJ...750L..43C}) consistent with previous spectroscopic RM results.

A fundamental issue in both these methods is to separate the host-galaxy
contribution from the total luminosity to calculate the AGN luminosity. 
An incorrect determination of the nuclear 
AGN luminosity (due to the contamination of the host component) causes an 
overestimation of the linear regression slope $\alpha$. This has been demonstrated 
by \citet{2009ApJ...697..160B}, who presented the most recent compilation of host-subtracted AGN
luminosities for several reverberation-mapped Seyfert 1 galaxies obtained through host-galaxy 
modeling of high-resolution $Hubble$ Space Telescope (HST) images. 
Compared to the previous slope of $\alpha = 0.7$ they determined 
an improved slope of $\alpha =$ 0.519$^{0.063}_{-0.066}$ close to $\alpha = 0.5$ expected 
from photo-ionization models of the BLR (\citealt{1979RvMP...51..715D}). 
A different approach to estimate the host galaxy contribution is the flux 
variation gradient method (FVG, \citealt{1981AcA....31..293C}; \citealt{1997MNRAS.292..273W}). An advantage is that it does not require
high spatial resolution imaging and can be applied directly to the monitoring data.
Recently, FVG has been tested on PRM data 
(\citealt{2011A&A...535A..73H}; \citealt{2012A&A...545A..84P}). These tests show that by using a 
well defined range for the host galaxy slope (\citealt{2010ApJ...711..461S}) it is possible 
to separate the AGN contribution at the time of the monitoring campaigns.

ESO399-IG20 has been classified as a Narrow-Line Seyfert 1 (NLS1) galaxy 
(\citealt{2010A&A...518A..10V}). 
It has often been speculated that the lack of broad emission lines of -- at least -- part of the NLS1 
population can be explained by face-on disk-like BLR geometries.
Single-epoch spectroscopic investigations by \citet{2005ApJ...623..700D} show 
high-ionized gas orbiting at high velocities, producing strong broad 
profiles (FWHM(H$\beta$) = 2425$\pm$121 km/s, see their Table 3 and Figures 2-5) like in 
Broad-Line Seyfert 1 (BLS1) galaxies. Furthermore, they find a strong 
host-galaxy contribution with respect to the total observed 5100\AA~ continuum 
flux ($\sim$ 50\%).
Here, we present the first  measurement of the BLR size and the 
host-subtracted AGN 5100\AA~ luminosity for \object{ESO399-IG20}. Both results allow us 
to infer its position in the BLR size-Luminosity $R_{\rm{BLR}} \propto L_{\rm{AGN}}^{0.5}$ diagram. 
Additionally, pure broad-band PRM (\citealt{2012ApJ...747...62C}) is applied 
and compared with the broad- and narrow-band PRM technique (\citealt{2011A&A...535A..73H}). 
Finally, the geometry of the BLR, whether spherical or an inclined disk, is inferred via 
light curve modeling.

\begin{figure}
  \centering
  \includegraphics[angle=0,width=\columnwidth]{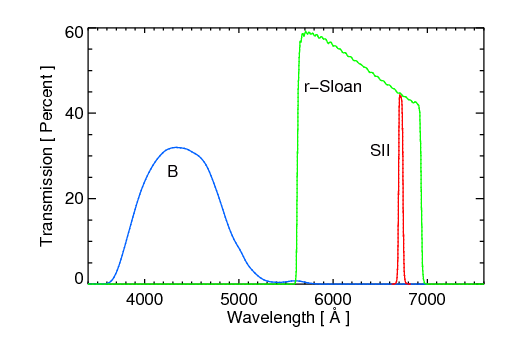}
  \includegraphics[angle=0,width=\columnwidth]{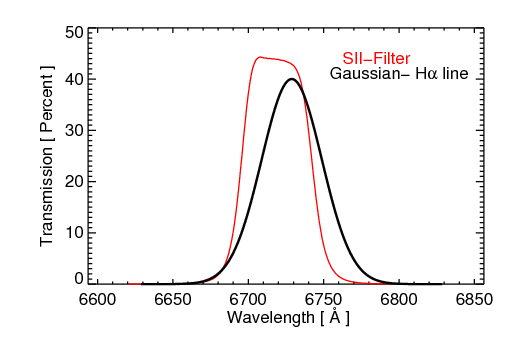}
  \caption{Effective transmission of the VYSOS6 filters convolved with the quantum efficiency of the ALTA U16M CCD camera (top).
    The position of the narrow-band [SII] 6721\AA~filter (red line) and the simulated redshifted gaussian-shaped H$\alpha$ emission line using the parameters obtained by \citet{2005ApJ...623..700D} (black line) are shown in the bottom panel. 
  }
  \label{filters}
\end{figure}

\section{Observations and data reduction}

\begin{figure*}
  \centering
  \includegraphics[width=18cm,clip=true]{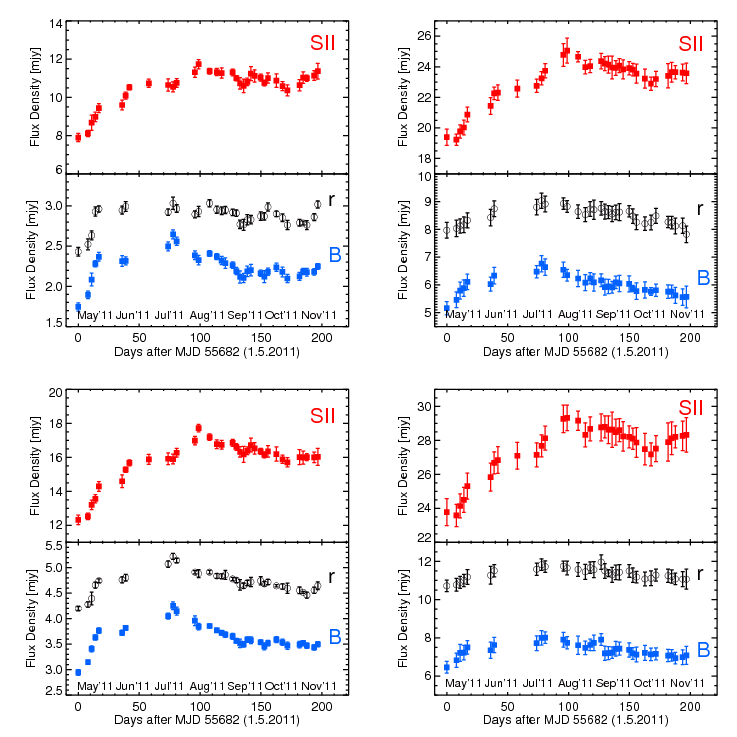}
  \caption{ 
    $B$,$r$-Sloan and SII-bands light curves for different apertures (5" upper left, 7.5" lower left, 15" upper right and 25" lower right). The light curves have been corrected for foreground galactic extinction using the values given in Table~\ref{table1}.}
  \label{first_lc}
\end{figure*}

Broad-band Johnson $B$ ($4330$\,\AA), Sloan-band $r$ ($6230$\,\AA) and the redshifted H$\alpha$ 
(SII $6721 \pm 30$\,\AA~ at z=0.0249) images were obtained with the robotic 15 cm VYSOS6 telescope of the
Universit\"atssternwarte Bochum, located near Cerro Armazones, future location of the ESO Extreme 
Large Telescope (ELT) in Chile. Monitoring occurred between May 5 and November 
18 of 2011, with a median sampling of 3 days.

ESO399-IG20 lies at redshift $z =0.0249$, therefore the H$\alpha$ emission line falls into 
the SII $6721 \pm 30\AA$ narrow-band filter. Figure~\ref{filters} shows the position of the narrow-band with 
respect to the H$\alpha$ emission line together with the effective transmission of the other filters used. 
The H$\alpha$ line is broader than the SII filter, however, simulations with different asymmetric and perfect symmetric 
cutting of the high-velocity line wings show negligible systematic effect introduced in the time delay ($\sim$2\%), even considering the fact of only using one part of the wing profile. This effect will be discussed in depth in a forthcoming contribution (Bruckmann et al. 2012., in preparation).

Data reduction was performed using IRAF\footnote{IRAF is distributed by the National Optical Astronomy Observatory, 
which is operated by the Association of Universities for Research in Astronomy (AURA) under cooperative agreement with the National Science Foundation.},
in combination with SCAMP (\citealt{2006ASPC..351..112B}) and SWARP (\citealt{2002ASPC..281..228B}) routines, in the same manner as described by \citet{2012AN....333..706H}.
Light curves were extracted using different apertures (5", 7$\farcs$5, 15" and 25") in order to compare and trace the host and AGN contribution. The light curves for the different apertures are shown in Figure~\ref{first_lc}. Additionally, for each aperture we determined the level of variability using the fractional flux variation introduced by \citet{1997ApJS..110....9R}:
\begin{equation}
{F_{\rm var}} = {\frac{\sqrt{\sigma^{2}-\Delta{^2}}}{\langle f\rangle}}
\end{equation}
where $\sigma^{2}$ is the flux variance of the observations, $\Delta^{2}$ is the mean square uncertainty, and $\langle f\rangle$ is the mean observed flux. The variability statistics is listed in Table~\ref{table2}. 
Selecting a small aperture (5")
we are only considering a small portion of the total flux, which remains heavily dependent on the quality of the PSF, whereas bigger apertures (15" and 25"), 
enclose more contribution of the host-galaxy and the results
are more sensitive to sky background contamination resulting in larger uncertainties. We found that 7$\farcs$5 around the nucleus is the right size aperture 
which maximizes the signal-to-noise ratio (S/N) and delivers the lowest absolute scatter for the fluxes. Thus, we use this aperture for the further analysis of the BLR size and AGN luminosity. Although the light curves in Figure~\ref{first_lc} have not yet been corrected for the host galaxy contribution, one can see that the fractional variation ($F_{var}$) and the ratio of the maximum to minimum fluxes ($R_{max}$) are higher in the $B$-band than in the $r$-band for which the host-galaxy present a greater contribution. A different case can be seen in the SII light curves, which are mostly dominated by the contribution of the H$\alpha$ emission line resulting in a higher amplitude of variability. As expected, for bigger apertures the $B$ and $r$ light curves become flatter with $F_{var}$ and $R_{max}$ smaller due to the larger contribution from the host-galaxy.

Non-variable reference stars located on the same field and with similar brightness as the AGN were used to create the relative light curves in normalized flux units. For the absolute photometry calibration we used reference stars from \citet{2009AJ....137.4186L} observed on the same nights as the AGN, considering the atmospheric extinction for the nearby site Paranal by \cite{2011A&A...527A..91P} and the recalibrated galactic foreground extinction presented by \citet{2011ApJ...737..103S} obtained from the \citet{1998ApJ...500..525S} dust extinction maps. In Table~\ref{table1}, we give the characteristics of ESO399-IG20. A summary of the
photometric results and the fluxes in all bands are listed in Table~\ref{table3} and Table~\ref{table4}, respectively.

\begin{table*}
\begin{center}
\caption{Characteristics of \object{ESO399-IG20}}
\label{table1}
\begin{tabular}{@{}cccccccc}
\hline\hline
$\alpha$ (2000)$^{(1)}$ & $\delta$ (2000)$^{(1)}$ & z$^{(1)}$ & $D_L^{(1)}$ & $B-V^{(2)}$ & $M_{abs}^{(2)}$ & $A_B^{(3)}$ & 
$A_V^{(3)}$ \\
      & & & (Mpc) & (mag) & (mag) & (mag) & (mag) \\
\hline
20:06:57.7 & -34:32:58.0 & 0.024951 & 102.0 & 0.96 & -19.7 & 0.409 & 0.309 \\
\hline
\end{tabular}
\end{center}
\tablefoottext{1}{Values from NED database,} 
\tablefoottext{2}{\cite{2010A&A...518A..10V},}
\tablefoottext{3}{\cite{2011ApJ...737..103S}.}
\end{table*}

\begin{table*}
\begin{center}
\caption{Light Curve Variability Statistics.}
\label{table2}
\begin{tabular}{lcccccccccccccc}
\hline\hline
& & $B$-band & & & & $r$-band & & & & SII-band &&\\
\hline\hline
Aperture Diameter & $<f>$ & $\sigma$ & $F_{var}$ & $R_{max}$ & $<f>$ & $\sigma$ & $F_{var}$ & $R_{max}$ & $<f>$ & $\sigma$ & $F_{var}$ & $R_{max}$ &&\\
(1) & (2) & (3) & (4) & (5) & (6) & (7) & (8) & (9) & (10) & (11) & (12) & (13) \\
\hline
5.0 & 2.24  &0.03  & 0.02  & 1.52 & 2.86 & 0.02 & 0.01 & 1.25 & 10.56 & 0.85 & 0.08 & 1.49 &&\\
7.5 & 3.63  &0.06  & 0.03  & 1.45 & 4.71 & 0.04 & 0.01 & 1.24 & 15.80 & 1.55 & 0.10 & 1.44 &&\\
15.0 & 5.98 &0.12  & 0.01  & 1.31 & 8.47 & 0.09 & 0.01 & 1.15 & 23.06 & 2.33 & 0.09 & 1.30 &&\\
25.0 & 7.35 &0.14  & 0.01  & 1.24 & 11.3 & 0.09 & 0.01 & 1.11 & 27.51 & 2.42 & 0.08 & 1.24 &&\\
\hline
\end{tabular}
\end{center}
Note. Column 1 list the aperture in arc-seconds, Columns 2-5 list the mean flux, the standard deviation, the normalized excess variance and the ratio of the maximum to minimum fluxes for the continuum light curve. Similarly, columns 6-13 list the variability for the $r$ and SII bands respectively. The standard deviation and the mean fluxes are expressed in units of mJy.\\
\end{table*}

\begin{table*}
\begin{center}
\caption{Summary of the photometry results for 7$\farcs$5 aperture.}
\label{table3}
\begin{tabular}{@{}cccccccc}
\hline\hline
$B$ & $r$ & SII & $fB_{total}^{(1)}$ & $fr_{total}^{(1)}$& $f$SII$_{total}^{(1)}$ \\
(mag) & (mag) & (mag) & (mJy) & (mJy) & (mJy) \\
\hline
15.05-15.46 & 14.57-14.81 & 13.21-13.58 & 3.45$\pm$0.06 & 4.87$\pm$0.08 & 17.17$\pm$0.34 \\
\hline
\end{tabular}
\end{center}
\tablefoottext{1}{$fB_{total}$ , $fr_{total}$ and $f$SII$_{total}$ refer to the mean of the total flux ranges during our monitoring.}
\end{table*}

\section{Results and discussion}

\subsection{Light curves and BLR size}

The $B$-band, which is dominated by the AGN continuum, shows a strong flux increase (about 35\%) between the beginning and the end of May.
Afterwards, the two measurements obtained in June reflect a more gradual increase until a maximum is reached at the end of July. After this maximum,
the fluxes decreases gradually (about 20\%) until the end of September, and then the light curves become more constant until middle of November. Likewise, 
the $r$-band, which is dominated by the continuum but also contains a contribution from the strong H$\alpha$ emission line, follows the sames features
as the $B$-band light curve, although with a lower amplitude. 
In contrast to the continuum dominated broad-band light curves, 
the narrow
SII-band light curve exhibits a flux increase which is stretched. 
For instance the prominent maximum in the $B$- and $r$- band at the end of July occurs in the SII-band in August, 
giving a first approximation for the H$\alpha$ time delay of 15-20 days.  
Note that the visual inspection of the $r$- and $B$-band light curves does not allow an approximation for the time delay.

\begin{figure}
  \centering
  \includegraphics[width=\columnwidth]{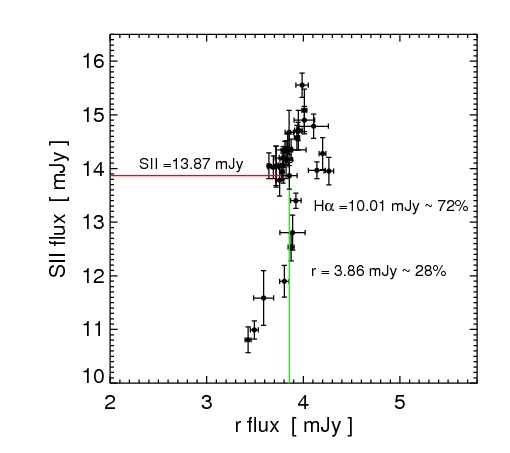}
  \caption{
    Flux-flux diagram for the SII and $r$ band measured using 7.5" aperture.
    Black dots denote the measurement pair of each night.
    The straight red and green lines represents the average flux in 
    the SII and $r$ band respectively.
    The data are as observed and not corrected for galactic foreground extinction. 
  }
\label{frac_ESO399}
\end{figure}

\begin{figure}
  \centering
  \includegraphics[width=\columnwidth]{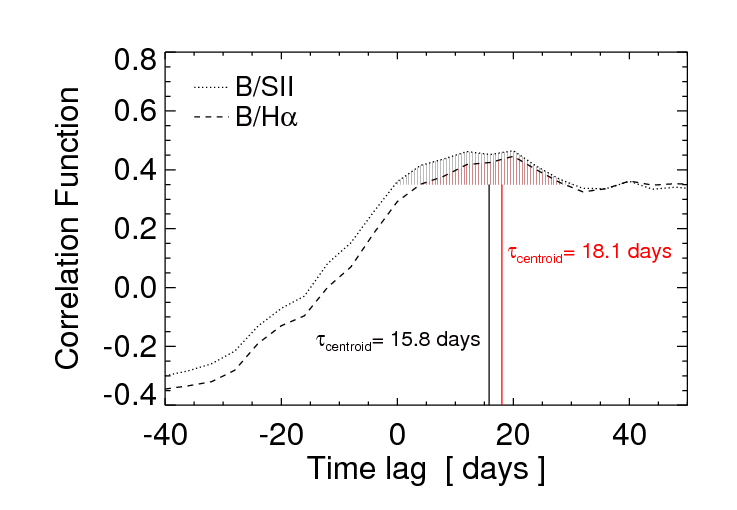}
  \caption{ 
    Cross  correlation of $B$ and SII
    light  curves (dotted line) and  of $B$ and H$\alpha$
    light  curves (dashed line). 
    The error  range ($  \pm
    1\sigma$) around the cross correlation was omitted for better
    viewing. The red and black shaded areas marks the range
    used to calculate the centroid of the lag 
    (vertical red and black straight lines).
  }
\label{DCF_ESO399}
\end{figure}

\begin{figure}
  \centering
  \includegraphics[height=7.5cm,width=\columnwidth]{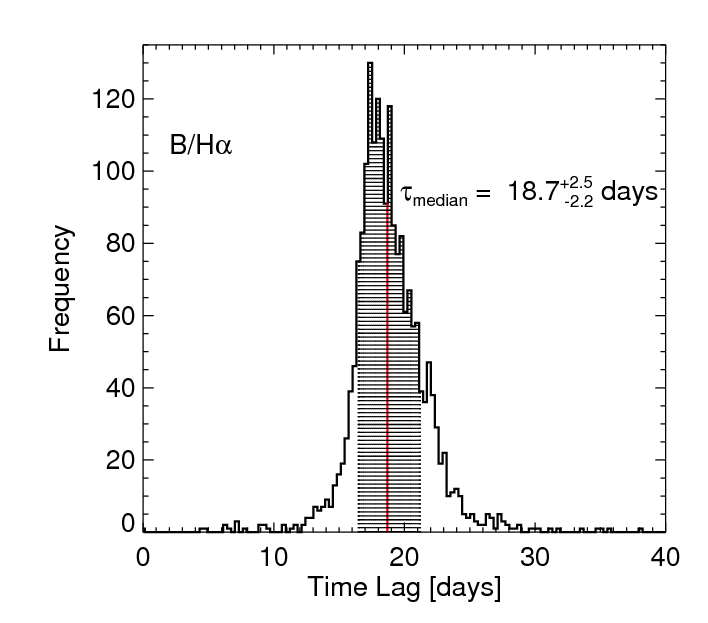}
  \caption{ Results of the lag error analysis. The histogram shows the
    distribution of the centroid lag obtained by cross correlating 2000
    flux randomized and randomly selected subset light curves 
    (FR/RSS method). The black
    area marks the 68\% confidence range used to calculate the errors of
    the centroid (red line).}
  \label{FR_RSS_ESO399}
\end{figure}

\begin{table*}
\begin{center}
\caption{$B$, $r$, $SII$ and H$\alpha$ Fluxes corrected by extinction.}
\label{table4}
\begin{tabular}{ccccc}
\hline
JD-2,450,000 & $F_{B}$ & $F_{r}$ & $F_{SII}$ & $F_{H\alpha}$ \\
             & (mJy)   &  (mJy)  & (mJy) & (mJy)  \\
\hline
\hline
55687.340 &  $2.944\pm 0.058$ & $4.199\pm  0.038$ & $12.318\pm 0.274$ &  $11.058\pm  0.276$ \\
55695.316 &  $3.147\pm 0.036$ & $4.279\pm  0.052$ & $12.528\pm 0.194$ &  $11.244\pm  0.200$ \\
55698.332 &  $3.407\pm 0.058$ & $4.396\pm  0.127$ & $13.206\pm 0.281$ &  $11.887\pm  0.308$ \\
55701.352 &  $3.631\pm 0.051$ & $4.659\pm  0.061$ & $13.561\pm 0.239$ &  $12.163\pm  0.246$ \\
55704.281 &  $3.765\pm 0.054$ & $4.744\pm  0.038$ & $14.287\pm 0.291$ &  $12.863\pm  0.293$ \\
55723.359 &  $3.722\pm 0.051$ & $4.763\pm  0.060$ & $14.594\pm 0.371$ &  $13.165\pm  0.375$ \\
55726.305 &  $3.819\pm 0.047$ & $4.805\pm  0.066$ & $15.272\pm 0.161$ &  $13.830\pm  0.174$ \\
55729.289 &  $-    \pm -    $ & $-    \pm  -    $ & $15.676\pm 0.178$ &  $15.676\pm  0.178$ \\
55745.234 &  $-    \pm -    $ & $-    \pm  -    $ & $15.886\pm 0.291$ &  $15.886\pm  0.291$ \\
55761.137 &  $4.051\pm 0.058$ & $5.073\pm  0.064$ & $15.918\pm 0.339$ &  $14.396\pm  0.344$ \\
55765.148 &  $4.253\pm 0.076$ & $5.223\pm  0.061$ & $15.902\pm 0.291$ &  $14.335\pm  0.297$ \\
55768.117 &  $4.149\pm 0.080$ & $5.143\pm  0.042$ & $16.273\pm 0.258$ &  $14.730\pm  0.261$ \\
55783.129 &  $3.960\pm 0.094$ & $4.913\pm  0.032$ & $16.983\pm 0.252$ &  $15.509\pm  0.254$ \\
55786.133 &  $3.845\pm 0.065$ & $4.885\pm  0.080$ & $17.726\pm 0.220$ &  $16.260\pm  0.234$ \\
55795.133 &  $3.856\pm 0.043$ & $4.913\pm  0.033$ & $17.193\pm 0.194$ &  $15.719\pm  0.196$ \\
55801.082 &  $3.772\pm 0.043$ & $4.838\pm  0.047$ & $16.774\pm 0.258$ &  $15.322\pm  0.262$ \\
55805.109 &  $3.722\pm 0.047$ & $4.833\pm  0.047$ & $16.741\pm 0.258$ &  $15.291\pm  0.262$ \\
55814.109 &  $3.686\pm 0.054$ & $4.861\pm  0.088$ & $16.854\pm 0.210$ &  $15.395\pm  0.227$ \\
55817.117 &  $3.653\pm 0.058$ & $4.780\pm  0.033$ & $16.612\pm 0.194$ &  $15.178\pm  0.196$ \\
55820.105 &  $3.566\pm 0.047$ & $4.755\pm  0.047$ & $16.322\pm 0.307$ &  $14.895\pm  0.310$ \\
55823.090 &  $3.516\pm 0.083$ & $4.664\pm  0.103$ & $16.176\pm 0.468$ &  $14.776\pm  0.479$ \\
55826.094 &  $3.494\pm 0.054$ & $4.645\pm  0.052$ & $16.386\pm 0.226$ &  $14.992\pm  0.231$ \\
55829.094 &  $3.588\pm 0.054$ & $4.701\pm  0.052$ & $16.725\pm 0.404$ &  $15.314\pm  0.407$ \\
55832.094 &  $3.577\pm 0.065$ & $4.720\pm  0.093$ & $16.531\pm 0.339$ &  $15.115\pm  0.351$ \\
55837.094 &  $3.537\pm 0.047$ & $4.744\pm  0.080$ & $16.354\pm 0.178$ &  $14.930\pm  0.195$ \\
55840.102 &  $3.458\pm 0.069$ & $4.683\pm  0.042$ & $16.144\pm 0.178$ &  $14.739\pm  0.182$ \\
55843.102 &  $3.519\pm 0.054$ & $4.720\pm  0.052$ & $16.354\pm 0.339$ &  $14.938\pm  0.342$ \\
55850.020 &  $3.588\pm 0.058$ & $4.645\pm  0.023$ & $16.192\pm 0.420$ &  $14.798\pm  0.420$ \\
55855.000 &  $3.537\pm 0.061$ & $4.631\pm  0.052$ & $15.886\pm 0.242$ &  $14.496\pm  0.247$ \\
55859.059 &  $3.469\pm 0.072$ & $4.598\pm  0.099$ & $15.708\pm 0.274$ &  $14.328\pm  0.291$ \\
55869.008 &  $3.490\pm 0.065$ & $4.556\pm  0.066$ & $16.015\pm 0.436$ &  $14.648\pm  0.440$ \\
55872.008 &  $3.519\pm 0.051$ & $4.518\pm  0.019$ & $15.983\pm 0.387$ &  $14.627\pm  0.387$ \\
55875.027 &  $3.472\pm 0.054$ & $4.462\pm  0.061$ & $16.015\pm 0.210$ &  $14.676\pm  0.218$ \\
55881.016 &  $3.436\pm 0.054$ & $4.556\pm  0.061$ & $15.999\pm 0.307$ &  $14.632\pm  0.313$ \\
55884.016 &  $3.498\pm 0.051$ & $4.645\pm  0.077$ & $16.031\pm 0.500$ &  $14.637\pm  0.505$ \\
\hline
\end{tabular}
\end{center}
\end{table*}

As already discussed in previous PRM studies, the narrow-band contains a contribution of the varying continuum, 
which must be removed before applying cross correlation techniques (\citealt{2011A&A...535A..73H}; \citealt{2012A&A...545A..84P}). In order to determine 
this contribution, we used the SII and $r$-band fluxes, previously calibrated to mJy, as is shown with the flux-flux diagram 
in Figure~\ref{frac_ESO399}. The H$\alpha$ line is strong contributing, on average, about 70\%  of the total flux enclose in 
the SII-band, while the continuum contribution ($r$-band) is about 30\%. Following the usual practice of PRM, we construct a 
synthetic H$\alpha$ light curve by subtracting a third of the $r$-band flux (H$\alpha$ = SII $-$ 0.3 $r$). The H$\alpha$
light curve was used afterwards to estimate
the time delay. For this purpose, we used the discrete correlation function (DCF, \citealt{1988ApJ...333..646E}) to cross-correlate the continuum 
and the synthetic H$\alpha$ emission line, taking into account possible bin size dependency (\citealt{1989A&A...219..101R}). The centroid in 
the cross-correlation of $B$-band and H$\alpha$ show a time delay of 18.1 days, while the centroid in the cross-correlation between $B$-band 
and SII-band yields a time delay of 15.8 days. However, this is an expected result because of the more pronounced peak at zero lag due to the contamination by the continuum emission in the narrow-band filter. Both cross-correlation functions are shown in Figure~\ref{DCF_ESO399}. Uncertainties in the time delay were calculated using the flux 
randomization and random subset selection method (FR/RSS, \citealt{1998PASP..110..660P}; \citealt{2004ApJ...613..682P}). From the observed light curves we create 
2000 randomly selected subset light curves, each containing 63\% of the original data points (the other fraction of points are unselected 
according to Poisson probability). The flux value of each data point was randomly altered consistent with its (normal-distributed) measurement error.
We calculated the DCF for the 2000 pairs of subset light curves and the corresponding centroid (Figure~\ref{FR_RSS_ESO399}). From this cross-correlation 
error analysis, we measure a median lag of $\tau_{cent}$ = 18.7 $^{+2.5}_{-2.2}$. Correcting for the time dilation factor ($1+z=1.0249$) 
we obtain a rest frame lag of $18.2 \pm 2.29$ days.

\begin{figure}
  \centering
  \includegraphics[width=\columnwidth]{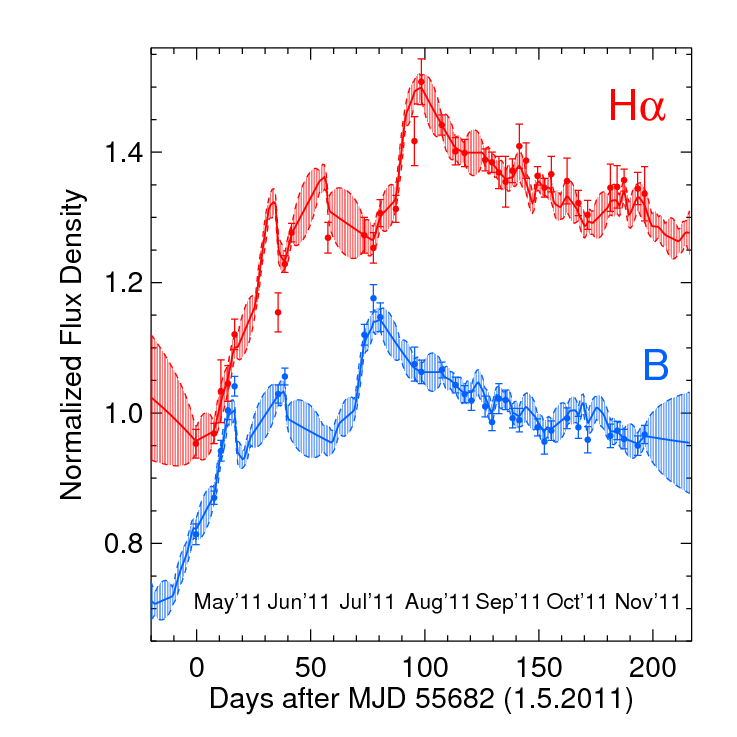}
  \caption{ 
    Synthetic H$\alpha$ and continuum light curves. The solid red and blue lines show the H$\alpha$
    and continuum models estimated by SPEAR respectively. The red and blue area (enclosed by the dashed line)
    represent the expected variance about the mean light curve model. The original H$\alpha$ light curve is computed by subtracting a scaled
    $r$ curve from the SII curve and re-normalizing it to mean $ = 1$ (red dots).
    The H$\alpha$ light curve is vertically shifted by 0.2 with respect to the continuum light curve (blue dots) for clarity.}
\label{lc_ESO399}
\end{figure}

An alternative method to estimate the time delay, called Stochastic Process Estimation 
for AGN Reverberation (SPEAR), has been worked out recently by \citet{2011ApJ...735...80Z}. Through the modeling of the AGN light curves as a damped 
random walk (DRW) (\citealt{2011ApJ...735...80Z}; \citealt{2012arXiv1202.3783Z} and reference therein), this method appears to be consistent with previous cross-correlation techniques 
(e.g. \citet{2012arXiv1206.6523G} for spectroscopic reverberation data). Using the SPEAR 
code\footnote{http://www.astronomy.ohio-state.edu/~yingzu/codes.html.} on our light curves, we obtain a time 
delay of $\tau_{spear}$ = 18.4 $^{+1.2}_{-1.0}$. After correction for time dilation the 
rest frame lag $\tau_{rest}$ = 17.9 $\pm 1.1$ days, in agreement  with the results from 
the DCF method. Figure~\ref{lc_ESO399} shows the SPEAR light curve models.

\subsection{Broad-band reverberation mapping approach}
\label{section_broadband}

Recently, pure broad-band photometric RM has been introduced as an efficient alternative 
to measure the BLR size in quasars (\citealt{2012ApJ...747...62C}). In this method, 
the emission line is also measured using a broad-band filter, and the removal of the underlying continuum 
is performed in the correlation domain.
The centroid of the lag is 
obtained by the subtraction of the auto-correlation function ($ACF$) between the continuum
(represented by one broad band filter enclosing only continuum emission) from the cross-correlation function ($CCF$)
between the continuum and the emission line (represented by one broad-band filter that contains a sufficiently strong emission line contribution).
This method has been applied for two objects; the low-luminosity AGN NGC4395 (\citealt{2012ApJ...756...73E}) and the high-redshift ($z=1.72$) 
luminous MACHO quasar (13.6805.324) (\citealt{2012ApJ...750L..43C}); in both cases a successful recovery of 
the time delay has been reported. 
Our \object{ESO399-IG20} has a strong H$\alpha$ emission line contributing to about 20\% to the $r$-band 
flux (Fig.\,\ref{frac_ESO399}) and the light curves are well sampled.
This makes \object{ESO399-IG20} an ideal object to perform a further test of the pure broad-band PRM method.

\begin{figure}
  \centering
  \includegraphics[width=10cm,clip=true]{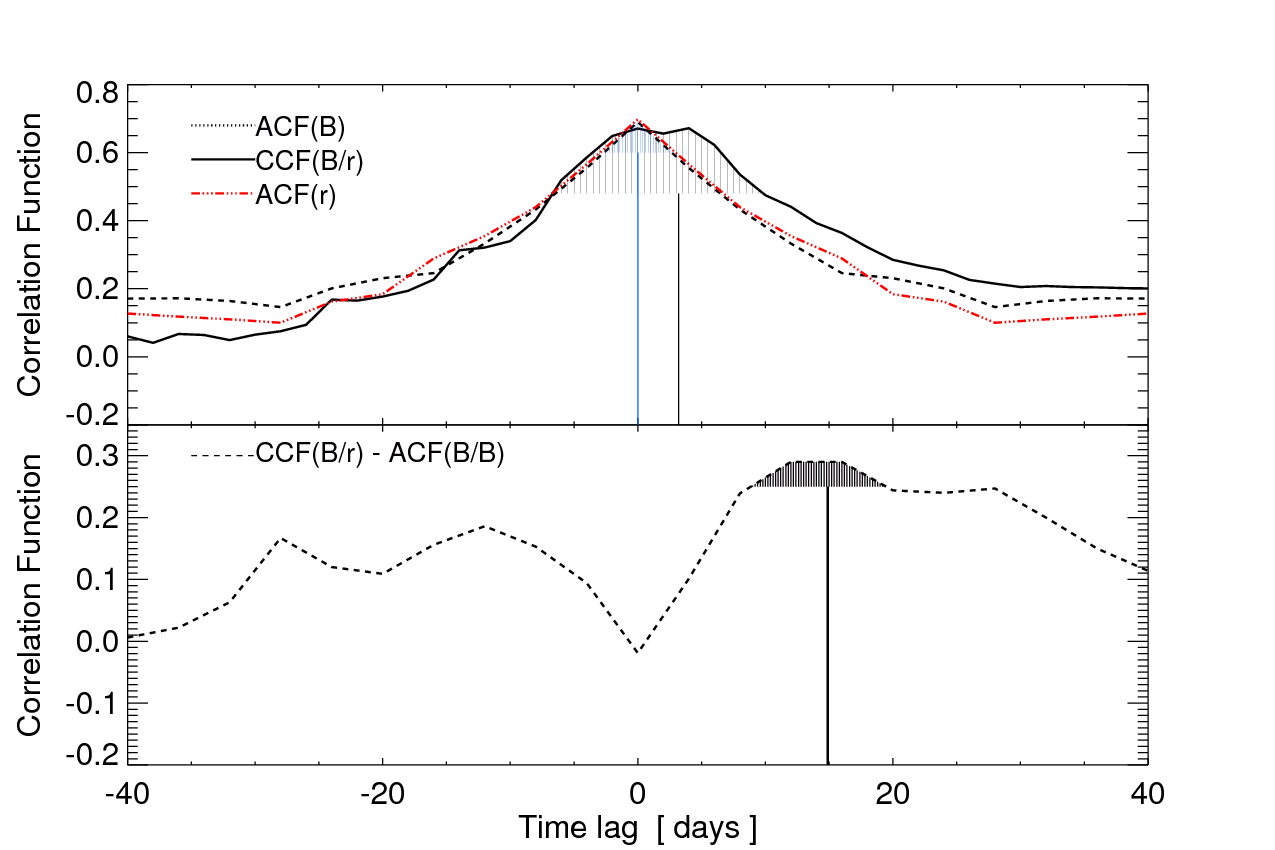}
  \caption{ 
    Broad-band RM results. Top: The $B$-band ACF (dotted line) show a well defined centroid at 
    zero lag (vertical blue line), while the $B/r$-band CCF shows the same peak at zero lag, a secondary peak at 
    3 days is also visible (vertical black line). In the same case as the $B$-band ACF, the $r$-band ACF (red line) show a peak at zero
    lag, however, one additionally faint peak is clearly visible between 10-20 days, in comparison with the ACF of the $B$-band. Bottom: The cross-correlation
    yields a centroid lag of 15.0 days (vertical black line). The red shaded area marks the range used to calculate the centroid.   
  }
\label{CCF_chelouche}
\end{figure}

\begin{figure}
  \centering
  \includegraphics[height=7.5cm,width=\columnwidth]{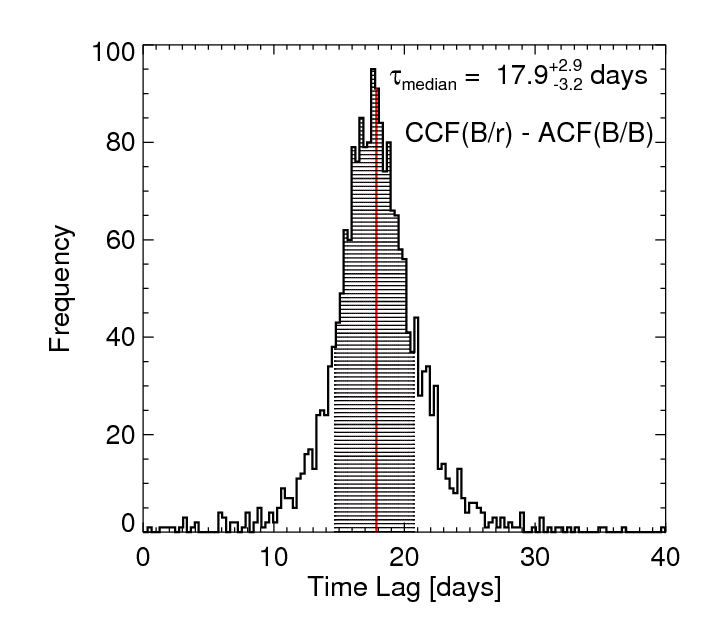}
  \caption{ Results of the lag error analysis of $CCF_{Br}(\tau) - ACF_{B}(\tau)$. 
    The histogram shows the
    distribution of the centroid lag obtained by cross correlating 2000
    flux randomized and randomly selected subset light curves 
    (FR/RSS method). The black shaded
    area marks the 68\% confidence range used to calculate the errors of
    the centroid (red line).}
  \label{FR_RSS_ESO399_chelouche}
\end{figure}

The $B$-band is mostly dominated by the continuum, and the $r$-band contains a strong H$\alpha$ emission line.
This allows
us to calculate the line-continuum cross-correlation function $CCF(\tau) = CCF_{Br}(\tau) - ACF_{B}(\tau)$ and 
to obtain the time delay (centroid). As shown in Figure~\ref{CCF_chelouche}, the $B/r$ cross-correlation exhibits two 
different peaks, one peak at zero lag (from the auto-correlation of the continuum) and one peak at lag $\sim$3 days. 
Furthermore, a small but 
extended enhancement can be seen in the auto-correlation of the $r$-band at about 10-20 days, in comparison with the ACF of the $B$-band. This feature can be interpreted as 
a clear contribution from the H$\alpha$ emission line to the broad $r$-band filter. In fact, the cross-correlation
shows a broad peak with a lag of 15.0 days as defined by the centroid in Figure~\ref{CCF_chelouche}. To determine the lag 
uncertainties, we applied the FR/RSS method. Again from the observed light curves we created 2000 randomly selected subset light curves, each containing 63\% of the original
data points, and randomly altering the flux value of each data point consistent with its (normal-distributed) measurement error. We calculated the DCF for the 2000 pairs
of subset light curves and the corresponding centroid (Figure~\ref{FR_RSS_ESO399_chelouche}). This yields a median lag $\tau_{cent}$ = 17.9 $^{+2.9}_{-3.2}$
days. Correcting for time dilation we obtain a rest frame lag $\tau_{rest}$ = 17.5 $\pm 3.1$ days.
This nicely agrees with the results from narrow-band PRM ($18.2 \pm 2.3$ days for the DCF, and 17.9 $\pm 1.1$ days for SPEAR)

\subsection{Modeling the geometry of the Broad-Line Region}
\label{section_geom_BLR}

\begin{figure}
  \centering
  \includegraphics[width=10cm,clip=true]{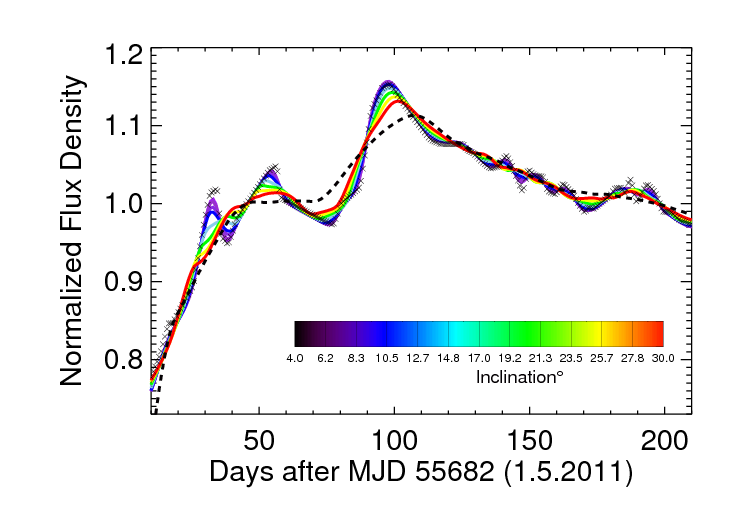}
  \caption{ 
    BLR disk/sphere-model. Simulated H$\alpha$ light curves for a disk-like BLR geometry and different inclinations ($4 \leq i \leq 30^\circ $) are shown in different colors for illustration. The black dotted line is the H$\alpha$ simulated light curve for a spherical BLR model. A disk-like BLR model with inclination of $6^\circ$ is able to reproduce the features of the original H$\alpha$ light curve (black crosses).
  }
\label{model_lc}
\end{figure}

We have inferred the geometry of the BLR through the direct modeling of the results obtained from PRM. Following \cite{1991ApJ...379..586W},
the time delay function for a spherically symmetric geometry of the BLR is:
\begin{eqnarray}
{{\Delta(\tau)}} =  \frac {r} {c} (1 + \sin \phi \cdot \cos \theta)
\end{eqnarray}
where $r$ is the radius of the spherical shell, $\phi$ and $\theta$ the respectives angles in the spherical coordinates. While for a Keplerian 
ring/disk structure the time delay function is:
\begin{eqnarray}
{{\Delta(\tau)}} =  \frac {r} {c} (1 - \sin i \cdot \cos \phi)
\end{eqnarray}
where $r$ is the radius of the ring, $i$ is the inclination ($0 \leq i
\leq 90^\circ $) of the axis of the disk with respect to the observer
line of sight ($0$=face-on, $90$=edge-on) 
and $\phi$ is the azimuthal angle between a point on the disk and the
projection of the line of sight onto the disk. 
The observed continuum light curve (previously interpolated by SPEAR)
was convolved with the time-delay function for the respective disk and
sphere geometry to simulate the expected H$\alpha$ light curve. 
The results of the simulation are shown in Figure~\ref{model_lc}.  

The H$\alpha$ light curve for a spherical BLR configuration does not reproduce the observations,
however a nearly face-on disk-like BLR geometry with an inclination
between $4$ to $10^\circ$ provides an acceptable fit to the original
data. 
To estimate the disk inclination, we performed the $\chi^2$ minimizing
fitting procedure for a range of inclination angles
(Fig~\ref{model_lc}).  
The best fit yields a value of $i=6^{\circ}\pm3^{\circ}$ for a
disk-like BLR model with an extension from 16 to 20 light days.
For such a nearly face-on BLR the velocity of orbiting broad
line gas clouds will appear about a factor ten smaller, mimicking a NLS1.

\begin{figure}
  \centering
  \includegraphics[angle=0,width=\columnwidth]{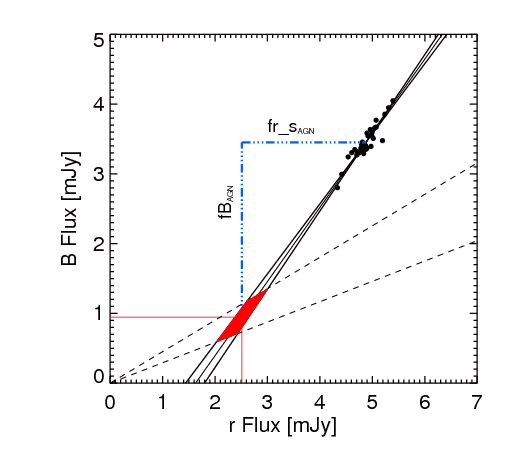} 
  \caption{Flux variation gradient diagram of \object{ESO399-IG20} for $7\farcs5$ aperture. 
    The solid lines delineate the
    ordinary least square bisector regression model yielding 
    the range of the AGN slope. The
    dashed lines indicate the interpolated range of host slopes 
    obtained from \citet{2010ApJ...711..461S} for 11 nearby AGN. 
    The intersection between the host galaxy and AGN slope (red area)
    gives the host galaxy flux in both bands. The dash-dotted
    blue lines represent the range of the AGN flux in both
    filters.
  }
  \label{fvg_first}
\end{figure}

\begin{table*}
\begin{center}
\caption{Total, host galaxy and AGN fluxes of ESO399-IG20}
\label{table5}
\begin{tabular}{@{}cccccccc}
\hline\hline
$fB_{total}$ & $fB_{host}$ & $fB_{AGN}^{1}$ & $fr_{total}$ & $fr_{host}$ & $fr_{AGN}^{1}$ & $f_{AGN} ((1+z)5100$\AA~ & $\lambda L_{\lambda,AGN}5100$\AA~ \\
(mJy) & (mJy)  &  (mJy) & (mJy) & (mJy) & (mJy) & ($10^{-15} erg s^{-1}cm^{-2}\AA^{-1}$) & ($10^{43} erg s^{-1}$) \\
\hline
3.45$\pm$0.11 & 0.95$\pm$0.18 & 2.50$\pm$0.21 & 4.87$\pm$0.10 & 2.51$\pm$0.20 & 2.36$\pm$0.22 & 2.66$\pm$0.21 & 1.69$\pm$0.25 \\
\hline
\end{tabular}
\end{center}
\tablefoottext{1}{AGN fluxes values $fB_{AGN} = fB_{total}-fB_{host}$ and $fr_{AGN}=fr_{total}-fr_{host}$, with uncertainty range $\sigma_{AGN} = (\sigma_{total}^{2} + \sigma_{host}^{2})^{0.5}$.}\\  
\end{table*}

\begin{figure}
  \centering
  \includegraphics[angle=0,width=\columnwidth]{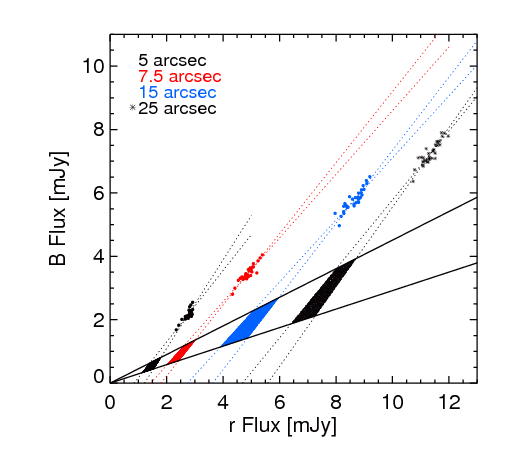} 
  \caption{Flux variation gradient diagram of \object{ESO399-IG20} for different apertures. The fluxes obtained through  $7\farcs5$ show an strong correlation, with a correlation coefficient of $r_{c}=0.94$. While for 5", 15" and 25" the correlation coefficients are equal to 0.81, 0.90 and 0.89 respectively. This significant decrease in the correlation supports once again our choice of $7\farcs5$ as the best aperture.
  }
  \label{fvg_plots}
\end{figure}

\subsection{AGN luminosity and the Host-subtraction process}
\label{section_agn_luminosity}

To determine the AGN luminosity free of host galaxy contributions, we
applied the flux variation gradient  
(FVG) method, originally proposed by \citet{1981AcA....31..293C} and later
modified by \citet{1992MNRAS.257..659W}.  
A detailed description of the FVG method on PRM data is presented in
\citet{2012A&A...545A..84P}, and here we give a brief outline. 
In this method the fluxes obtained through different filters and same
apertures are plotted in a flux-flux diagram. The fluxes follow a
linear slope representing the AGN color, while the slope of the
nuclear host galaxy contribution 
(including the contribution from the narrow line region (NLR)) lies in
a well defined range  ($0.4 < \Gamma^{host}_{BV} <
0.53$, for 8$\farcs$3 aperture and redshift $z<0.03$, \citealt{2010ApJ...711..461S}). 
The AGN slope is determined through linear
regression analysis. The intersection of the AGN slope with the host
galaxy range yields the actual host galaxy contribution at the time of
the monitoring campaign. Figure~\ref{fvg_first} shows the FVG
diagram for the $B$ and $r$ fluxes (corrected for galactic foreground
extinction) obtained during the same nights and through $7\farcs5$
aperture. Additionally, FVGs were evaluated for different apertures, as
shown in Figure~\ref{fvg_plots}. As already noted in Section 2, a big
fraction of the total flux is lost in the small 5" aperture,
hence, the results are more sensitive to a possible underestimation of
the real AGN and host galaxy contribution. Furthermore, one may expect
that for larger apertures the fluxes will lie closer to the range for
the host galaxy slope, however, it appears that the host galaxy is
intrinsically strong and very blue, closer to the nucleus and in the
outer parts. As for the analysis of the light curves we here use the
results for the 7$\farcs$5 aperture.

\begin{figure}
  \centering
  \includegraphics[width=\columnwidth,clip=true]{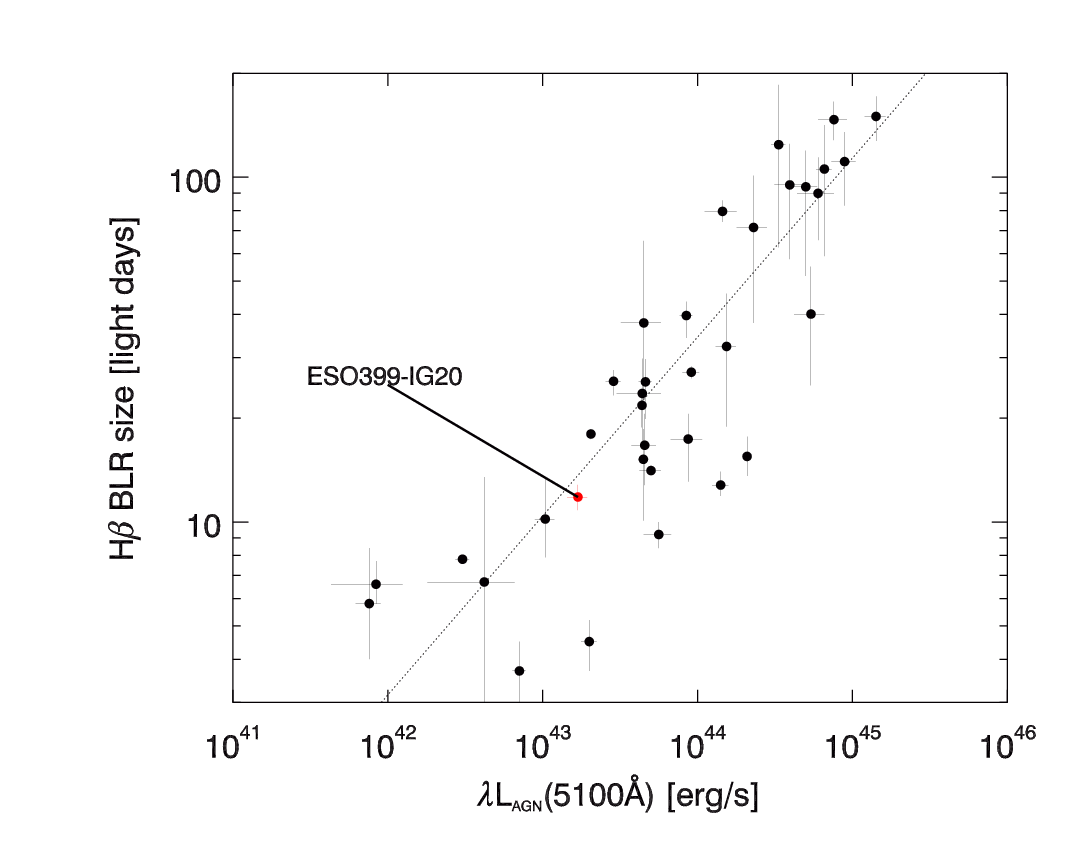}
  \caption{$R_{BLR}$ versus $L$, using data from \citet{2009ApJ...697..160B}, \citet{2010ApJ...721..715D}, \citet{2012MNRAS.426..416D} and \citet{2012arXiv1206.6523G}. As a red dot we included our results for ESO399-IG20. Shown is a zoomed portion containing ESO399-IG20. The solid dotted line show a fitted slope $\alpha = 0.519$. ESO399-IG20 lies close to the expected slope.}
  \label{BLR_luminosity}
\end{figure}

Averaging over the intersection area between the AGN and the host
galaxy slopes, we obtain a mean host galaxy flux of $(0.95 \pm 0.18)$
mJy in $B$ and $(2.51 \pm 0.20)$ mJy in $r$. During our monitoring campaign the 
host galaxy subtracted AGN fluxes range between 2.29 and 2.71 and between 
2.14 and 2.58 mJy in the $B$ and $r$ band, respectively. These
fluxes are represented by the blue dotted lines in
Figure~\ref{fvg_first}. From this range we interpolate the
host-subtracted monochromatic AGN flux at restframe $5100\AA~$
$F_{5100Å}=2.66 \pm 0.22 \cdot 10^{-15} erg s^{-1}cm^{-2}\AA^{-1}$. 
For the interpolation we assumed a power law spectral energy distribution (SED) ($F_{\nu} \propto
\nu^{\alpha}$) with an spectral index $\alpha=\log(fB_{AGN} / fr_{AGN}) / \log(\nu_{B} / \nu_{r})$,
where $\nu_{B}$ and $\nu_{r}$ are the effective frequencies in the $B$ and $r$ bands, respectively. 
The error was determined by interpolation between the ranges of the AGN fluxes $\pm\sigma$ in both filters. 
At the distance of 102 Mpc this yields a host-subtracted AGN luminosity at 5100\AA $L_{\rm{AGN}} = (1.69 \pm 
0.25)\times 10^{43}erg s^{-1}$. The total fluxes, host galaxy
subtracted AGN fluxes and the AGN luminosity are listed in
Table~\ref{table5}.

\subsection{The BLR size-luminosity relationship}
\label{section_agn_luminosity}

Estimates of the BLR size and host-galaxy subtracted AGN luminosity in the literature have been derived from several spectroscopic RM campaigns and through host-galaxy modeling using high-resolution images from $HST$. 
In consequence, the relationship between the H$\beta$ BLR size 
and the luminosity (5100\AA) $R_{{\rm BLR}} \propto L^{\alpha}$ (\citealt{2000ApJ...533..631K}) has been 
improved considerably with the most recent slope of $\alpha =$ 0.519$^{0.063}_{-0.066}$ (\citealt{2009ApJ...697..160B}). Although to date this relationship has been corroborated for 38 AGNs, still there exist objects with large uncertainties in both measurements. In order to improve the statistic, it is of interest to see the position for this 
new Seyfert 1 galaxy on the BLR-Luminosity relationship. In order to obtain the H$\beta$ BLR radius, we used the
weighted mean ratio for the time lag $\tau(H\alpha):\tau(H\beta):1.54:1.00$, obtained recently by \citet{2010ApJ...716..993B} from
the Lick AGN Monitoring Program of 11 low-luminosity AGN. Therefore,  the
H$\alpha$ lag of $18.2$ days translates into an H$\beta$ lag of $11.8$ days. Figure~\ref{BLR_luminosity} shows the position of ESO399-IG20 on the $R_{{\rm BLR}}$-$L_{AGN}$ diagram. The data are taken from \citet{2009ApJ...697..160B} and here we include the most recent results for particular objects obtained from spectroscopic RM by \citet{2010ApJ...721..715D}, \citet{2012MNRAS.426..416D} and \citet{2012arXiv1206.6523G} respectively.

\section{Summary and conclusions}
\label{section_conclusions}

We presented new photometric reverberation mapping results for
the Seyfert 1 galaxy ESO399-IG20. We determined 
the broad line region size, the basic geometry of the BLR and the
host-subtracted AGN luminosity. The results are: 

\begin{enumerate}
\item  The cross-correlation of the H$\alpha$ emission line measured
  in a  narrow-band filter 
  with the optical continuum light curve yields a rest-frame time
  delay $\tau_{rest}$ = 18.2 $\pm$ 2.29 days. 
  We explored the SPEAR method, and -- given that H$\alpha$ contributes
  about 15-20\% to the $r$-band -- also the capabilities of pure broad-band
  photometric reverberation mapping. 
  The SPEAR method
  yields a rest-frame time delay of $\tau_{rest}$ = 17.9 $\pm$ 1.1
  days, while the pure broad-band PRM method yields $\tau_{rest}$ = 17.5 $\pm$ 3.1
  days. The results indicate that, within the errors, the
  three methods are in good agreement.  

\item We constrained the basic geometry of the BLR by comparing simulated light curves, using the size 
determined for the BLR, to the observed H$\alpha$ light curve. 
  The pronounced and sharp variability features in both the continuum
  and emission line light curves allow us to exclude a spherical BLR
  geometry. 
  We found
  that the BLR has a disk-like shape with an inclination of
  $i=6\pm3^\circ$ and an extension from 16 to 20 light days. 
  This nearly face-on BLR can explain the appearance of ESO399-IG20
  as a narrow-line Seyfert-1 galaxy.
  
\item  We successfully separated the host-galaxy contribution from the
  total flux through the flux variation gradient method (FVG). The average
  host-galaxy subtracted AGN luminosity of \object{ESO399-IG20} at the
  time of our monitoring campaign is 
  $L_{\rm{AGN}} = (1.69 \pm 0.25)\times 10^{43}erg s^{-1}$. 
  In the BLR size -- AGN luminosity diagram ESO399-IG20 lies close to
  the best fit of the relation.
   
\end{enumerate}
These results document the efficiency and accuracy of photometric
reverberation mapping for determining the AGN luminosity, the BLR size
and the potential to constrain even the BLR geometry.

\begin{acknowledgements}
 
  This publication is supported as a project of the
  Nordrhein-Westf\"alische Akademie der Wissenschaften und der K\"unste
  in the framework of the academy program by the Federal Republic of
  Germany and the state Nordrhein-Westfalen.

  The observations on Cerro Armazones benefitted
  from the care of the guardians Hector Labra, Gerardo Pino, Roberto Munoz, 
  and Francisco Arraya.

  This research has made use of the NASA/IPAC
  Extragalactic Database (NED) which is operated by the Jet Propulsion
  Laboratory, California Institute of Technology, under contract with
  the National Aeronautics and Space Administration. This research has made 
  use of the SIMBAD database, operated at CDS, Strasbourg, France.
  We thank the anonymous referee for his comments and careful review of the manuscript.

\end{acknowledgements}

\bibliographystyle{aa} 
\bibliography{ESO_BLR}

\begin{thebibliography}{33}
\expandafter\ifx\csname natexlab\endcsname\relax\def\natexlab#1{#1}\fi

\bibitem[{{Bentz} {et~al.}(2009{\natexlab{a}}){Bentz}, {Peterson}, {Netzer},
  {Pogge}, \& {Vestergaard}}]{2009ApJ...697..160B}
{Bentz}, M.~C., {Peterson}, B.~M., {Netzer}, H., {Pogge}, R.~W., \&
  {Vestergaard}, M. 2009{\natexlab{a}}, \apj, 697, 160

\bibitem[Bentz et al.(2010)]{2010ApJ...716..993B} Bentz, M.~C., Walsh, 
J.~L., Barth, A.~J., et al.\ 2010, \apj, 716, 993


\bibitem[Bertin et al.(2002)]{2002ASPC..281..228B} Bertin, E., Mellier, Y., 
Radovich, M., et al.\ 2002, Astronomical Data Analysis Software and Systems 
XI, 281, 228 

\bibitem[Bertin(2006)]{2006ASPC..351..112B} Bertin, E.\ 2006, Astronomical 
Data Analysis Software and Systems XV, 351, 112 

\bibitem[{{Blandford} \& {McKee}(1982)}]{1982ApJ...255..419B}
{Blandford}, R.~D. \& {McKee}, C.~F. 1982, \apj, 255, 419

\bibitem[Chelouche 
\& Daniel(2012)]{2012ApJ...747...62C} Chelouche, D., \& Daniel, E.\ 2012, \apj, 747, 62

\bibitem[Chelouche et al.(2012)]{2012ApJ...750L..43C} Chelouche, D., 
Daniel, E., \& Kaspi, S.\ 2012, \apjl, 750, L43 

\bibitem[{{Choloniewski}(1981)}]{1981AcA....31..293C}
{Choloniewski}, J. 1981, \actaa, 31, 293

\bibitem[Davidson 
\& Netzer(1979)]{1979RvMP...51..715D} Davidson, K., \& Netzer, H.\ 1979, Reviews of Modern Physics, 51, 715 


\bibitem[Denney et al.(2010)]{2010ApJ...721..715D} Denney, K.~D., Peterson, 
B.~M., Pogge, R.~W., et al.\ 2010, \apj, 721, 715
 
\bibitem[Dietrich et al.(2005)]{2005ApJ...623..700D} Dietrich, M., 
Crenshaw, D.~M., \& Kraemer, S.~B.\ 2005, \apj, 623, 700 

\bibitem[Doroshenko et al.(2012)]{2012MNRAS.426..416D} Doroshenko, V.~T., 
Sergeev, S.~G., Klimanov, S.~A., Pronik, V.~I., 
\& Efimov, Y.~S.\ 2012, \mnras, 426, 416  


\bibitem[{{Edelson} \& {Krolik}(1988)}]{1988ApJ...333..646E}
{Edelson}, R.~A. \& {Krolik}, J.~H. 1988, \apj, 333, 646

\bibitem[Edri et al.(2012)]{2012ApJ...756...73E} Edri, H., Rafter, S.~E., 
Chelouche, D., Kaspi, S., \& Behar, E.\ 2012, \apj, 756, 73 

 
\bibitem[Grier et al.(2012)]{2012arXiv1206.6523G} Grier, C.~J., Peterson, 
B.~M., Pogge, R.~W., et al.\ 2012, arXiv:1206.6523


\bibitem[{{Haas} {et~al.}(2011){Haas}, {Chini}, {Ramolla}, {Pozo Nu{\~n}ez},
  {Westhues}, {Watermann}, {Hoffmeister}, \& {Murphy}}]{2011A&A...535A..73H}
{Haas}, M., {Chini}, R., {Ramolla}, M., {et~al.} 2011, \aap, 535, A73

\bibitem[Haas et al.(2012)]{2012AN....333..706H} Haas, M., Hackstein, M., 
Ramolla, M., et al.\ 2012, Astronomische Nachrichten, 333, 706 

\bibitem[Kaspi et al.(1996)]{1996ApJ...471L..75K} Kaspi, S., Smith, P.~S., 
Maoz, D., Netzer, H., \& Jannuzi, B.~T.\ 1996, \apjl, 471, L75 

\bibitem[{{Kaspi} {et~al.}(2000){Kaspi}, {Smith}, {Netzer}, {Maoz}, {Jannuzi},
  \& {Giveon}}]{2000ApJ...533..631K}
{Kaspi}, S., {Smith}, P.~S., {Netzer}, H., {et~al.} 2000, \apj, 533, 631

\bibitem[Kaspi et al.(2007)]{2007ApJ...659..997K} Kaspi, S., Brandt, W.~N., 
Maoz, D., et al.\ 2007, \apj, 659, 997 

\bibitem[{{Koratkar} \& {Gaskell}(1991)}]{1991ApJ...370L..61K}
{Koratkar}, A.~P. \& {Gaskell}, C.~M. 1991, \apjl, 370, L61

\bibitem[{{Landolt}(2009)}]{2009AJ....137.4186L}
{Landolt}, A.~U. 2009, \aj, 137, 4186

\bibitem[McGill et al.(2008)]{2008ApJ...673..703M} McGill, K.~L., Woo, 
J.-H., Treu, T., \& Malkan, M.~A.\ 2008, \apj, 673, 703 

\bibitem[Netzer 
\& Peterson(1997)]{1997ASSL..218...85N} Netzer, H., \& Peterson, B.~M.\ 1997, Astronomical Time Series, 218, 85 

\bibitem[Patat et 
al.(2011)]{2011A&A...527A..91P} Patat, F., Moehler, S., O'Brien, K., et al.\ 2011, \aap, 527, A91 


\bibitem[{{Peterson} {et~al.}(2004){Peterson}, {Ferrarese}, {Gilbert}, {Kaspi},
  {Malkan}, {Maoz}, {Merritt}, {Netzer}, {Onken}, {Pogge}, {Vestergaard}, \&
  {Wandel}}]{2004ApJ...613..682P}
{Peterson}, B.~M., {Ferrarese}, L., {Gilbert}, K.~M., {et~al.} 2004, \apj, 613,
  682

\bibitem[{{Peterson} {et~al.}(1998b){Peterson}, {Wanders}, {Horne}, {Collier},
  {Alexander}, {Kaspi}, \& {Maoz}}]{1998PASP..110..660P}
{Peterson}, B.~M., {Wanders}, I., {Horne}, K., {et~al.} 1998b, \pasp, 110, 660

\bibitem[Peterson 
\& Wandel(1999)]{1999ApJ...521L..95P} Peterson, B.~M., \& Wandel, A.\ 1999, \apjl, 521, L95 

\bibitem[Peterson(1993)]{1993PASP..105..247P} Peterson, B.~M.\ 1993, \pasp, 
105, 247 

\bibitem[Pozo Nu{\~n}ez et 
al.(2012)]{2012A&A...545A..84P} Pozo Nu{\~n}ez, F., Ramolla, M., Westhues, C., et al.\ 2012, \aap, 545, A84 

\bibitem[Rodriguez-Pascual et 
al.(1989)]{1989A&A...219..101R} Rodriguez-Pascual, P.~M., Santos-Lleo, M., \& Clavel, J.\ 1989, \aap, 219, 101

\bibitem[Rodriguez-Pascual et al.(1997)]{1997ApJS..110....9R} 
Rodriguez-Pascual, P.~M., Alloin, D., Clavel, J., et al.\ 1997, \apjs, 110, 
9 

\bibitem[{{Sakata} {et~al.}(2010){Sakata}, {Minezaki}, {Yoshii}, {Kobayashi},
  {Koshida}, {Aoki}, {Enya}, {Tomita}, {Suganuma}, {Katsuno Uchimoto}, \&
  {Sugawara}}]{2010ApJ...711..461S}
{Sakata}, Y., {Minezaki}, T., {Yoshii}, Y., {et~al.} 2010, \apj, 711, 461

\bibitem[Schlafly 
\& Finkbeiner(2011)]{2011ApJ...737..103S} Schlafly, E.~F., \& Finkbeiner, D.~P.\ 2011, \apj, 737, 103 

\bibitem[{{Schlegel} {et~al.}(1998){Schlegel}, {Finkbeiner}, \&
  {Davis}}]{1998ApJ...500..525S}
{Schlegel}, D.~J., {Finkbeiner}, D.~P., \& {Davis}, M. 1998, \apj, 500, 525

\bibitem[V{\'e}ron-Cetty 
\& V{\'e}ron(2010)]{2010A&A...518A..10V} V{\'e}ron-Cetty, M.-P., \& V{\'e}ron, P.\ 2010, \aap, 518, A10 

\bibitem[Vestergaard et al.(2011)]{2011nlsg.confE..38V} Vestergaard, M., 
Denney, K., Fan, X., et al.\ 2011, Narrow-Line Seyfert 1 Galaxies and their 
Place in the Universe,

\bibitem[Wandel et al.(1999)]{1999ApJ...526..579W} Wandel, A., Peterson, 
B.~M., \& Malkan, M.~A.\ 1999, \apj, 526, 579 

\bibitem[Welsh 
\& Horne(1991)]{1991ApJ...379..586W} Welsh, W.~F., \& Horne, K.\ 1991, \apj, 379, 586 

\bibitem[{{Winkler}(1997)}]{1997MNRAS.292..273W}
{Winkler}, H. 1997, \mnras, 292, 273

\bibitem[{{Winkler} {et~al.}(1992){Winkler}, {Glass}, {van Wyk}, {Marang},
  {Jones}, {Buckley}, \& {Sekiguchi}}]{1992MNRAS.257..659W}
{Winkler}, H., {Glass}, I.~S., {van Wyk}, F., {et~al.} 1992, \mnras, 257, 659

\bibitem[Zu et al.(2011)]{2011ApJ...735...80Z} Zu, Y., Kochanek, C.~S., 
\& Peterson, B.~M.\ 2011, \apj, 735, 80

\bibitem[Zu et al.(2012)]{2012arXiv1202.3783Z} Zu, Y., Kochanek, C.~S., 
Koz{\l}owski, S., \& Udalski, A.\ 2012, arXiv:1202.3783


\end{thebibliography}

\end{document}